\documentclass[12pt]{article}

\usepackage{sbc-template}
\usepackage{graphicx,url}
\usepackage[utf8, latin1]{inputenc}
\usepackage{wrapfig}

\title{Advancing Polyglot Big Data Processing \\using the Hadoop ecosystem}

\author{Antony Seabra, Sergio Lifschitz}

\address{Departamento de Informatica - PUC-Rio
  \email{\{amedeiros, sergio\}@inf.puc-rio.br}
}

\begin{document} 
\maketitle
\begin{abstract}
This article explores the utilization of the Hadoop ecosystem as a polyglot big data processing platform, focusing on the integration of diverse computation and storage technologies and their potential advantages in certain computational contexts. It delves into the potential of this ecosystem as a unified platform highlighting its architectural foundations and their complementary strengths in distributed storage, processing efficiency and real-time analytics. The article explores potential use cases  within domains such as Smart Cities and Social Networks, illustrating how the platform's diverse components can be orchestrated in a polyglot manner and how these fields can benefit from the ecosystem's capabilities. Finally, the article concludes by showcasing alternatives for future research, including specialized architectural aspects of the ecosystem to advance the polyglot paradigm.
\end{abstract}

\section{Introduction}
Processing large volumes of data has become a critical issue for organizations across all industries. With the proliferation of digital technologies and the advent of the internet era, businesses are inundated with vast amounts of data streaming in from various sources such as customer interactions, transactions, sensors, social media and more. Effectively harnessing this data holds immense importance for organizations as it provides valuable insights that can drive informed decision-making and strategic planning. From capturing  and storing information to processing and analyzing it, organizations face the challenge of managing the velocity, volume and variety of data - concepts often used to define the term Big Data.

Over the past two decades, the Hadoop ecosystem has evolved significantly from its initial use as a solution for distributed web page indexing to its current role as a comprehensive platform. This transformation has seen Hadoop become a foundational framework for constructing data lakes, adept at handling structured, semi-structured and unstructured data - one of the main reasons not to use traditional database management systems. The ecosystem initially relied on the technologies HDFS and MapReduce; however, with the integration of Spark, it substantially improved its processing prowess, notably in managing large and intricate data sets. Additionally, the evolution of various components, such as the data warehousing framework Hive and the NoSQL database HBase, has enhanced the ecosystem's proficiency in effectively managing and processing queries from a multitude of sources.

Big data analysis presents unique challenges due to its diverse nature and massive scale. One key aspect is the need to accommodate various file formats and processing characteristics. Unlike traditional data processing, where structured data in relational databases predominates, big data encompasses a wide array of data types and requires different approaches to load and to process the data. 

For instance, processing log files and IoT sensor data presents distinct challenges due to differences in data structure, volume and velocity. Log files are semi-structured and generated at a consistent rate, primarily serve monitoring and incident analysis purposes. In contrast, IoT sensor data is highly diverse, generated at a higher velocity, and often requires real-time processing to facilitate immediate decision-making or actions. Whereas the analysis of log files can typically be conducted in batches and is retrospective in nature, IoT data processing demands more dynamic, often edge-based, real-time analytics and storage solutions to handle the continuous and voluminous stream of data from various sensors.

In this work, we show that Polyglot Big Data Processing, a mixture of polyglot computing and polyglot persistence, is a versatile approach in big data analysis that uses multiple processing engines and data stores to efficiently handle different types of data and processing needs. In environments like Hadoop and Spark, it allows for the use of specialized building blocks, such as Apache Pig for data flow scripting, Apache HBase for NoSQL data storage and retrieval, and Apache Kafka for real-time data stream processing. The aim of this study is to discern the present components of the Hadoop ecosystem that can be used in conjunction to process large volumes of data, in a polyglot data processing approach. By confronting the tasks related to industry's needs with the specificities of the ecosystem components, we can conclude about which of them are more suitable for one or another task, including a comparative analysis of these polyglot strategies against traditional, or monoglot, systems, thereby highlighting the relative advantages and limitations of utilizing the Hadoop ecosystem in contrast to conventional data processing systems.

The paper is structured as follows. Section 2 presents an overview of the Hadoop ecosystem, detailing its core components and their roles in big data processing. Section 3 explores the concept of Polyglot Persistence and the types of Polyglot Data Stores. Section 4 delves into the Polyglot nature of the Hadoop ecosystem, discussing how it supports multiple data processing paradigms and storage systems to accommodate different data processing needs. Section 5 presents use cases related to Polyglot Big Data Processing implemented on the Hadoop ecosystem, illustrating practical applications and benefits of this approach. Related work is presented in Section 6. Finally, our conclusions and directions for future work are presented in Section 7.

\section{Overview of the Hadoop ecosystem}
Hadoop is an open-source platform for distributed storage and processing of large-scale datasets, inspired by the Google File System (GFS) [19] and MapReduce [14] papers published by Google. Hadoop was created by Doug Cutting and Mike Cafarella in 2005, while they were working at Yahoo. The name Hadoop comes from a toy elephant owned by Cutting's son, which also served as the inspiration for the project's logo.

The main motivation for Hadoop was to address the challenges of processing and analyzing large-scale datasets that were too big to fit on a single machine's storage or memory. Initially, Hadoop consisted of two main components: Hadoop Distributed File System (HDFS) for distributed storage, and MapReduce for distributed processing. Over the years, Hadoop has evolved significantly, with the addition of new components, such as Yarn, which became the default cluster management tool in Hadoop 2.0, and a range of complementary tools and frameworks [16]. Nowadays, the Hadoop ecosystem has become a critical part of the big data technology stack, enabling organizations to store, process and conduct data-oriented analysis of large-scale datasets that facilitate the generation of insights and the realization of business value. Figure 1 shows some of the most popular building blocks of this ecosystem.

\begin{figure}[ht]
\centering
\includegraphics[width=.70\textwidth]
{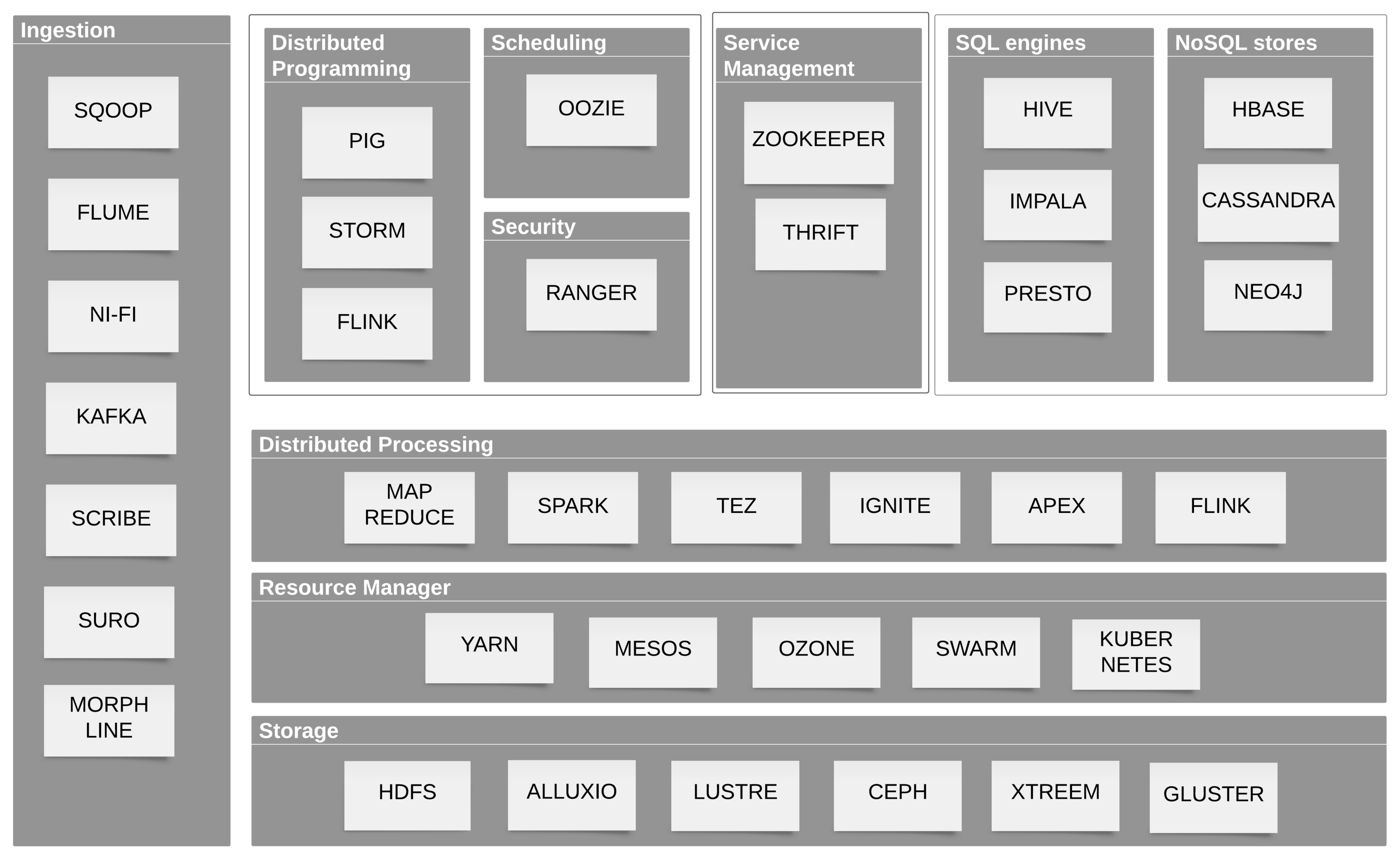}
\caption{Hadoop ecosystem}
\label{fig:hadoop }
\end{figure}

In addition, Hadoop's open-source nature allowed for widespread adoption, leading to a vibrant and active community of developers contributing to its ongoing development and evolution. The Apache Software Foundation, which stewards the Hadoop project, plays a pivotal role in coordinating the development efforts and ensuring that Hadoop remains a high-quality, reliable framework for big data processing. Major technology companies such as Facebook (now Meta), Twitter and Netflix have also been instrumental in Hadoop's growth, not only by using the framework at scale within their operations but also by contributing code, sharing use cases and participating in the ecosystem's expansion. Facebook, for instance, has developed and contributed several projects that integrate with Hadoop, like [7], while Twitter and Netflix have shared experiences, tools and libraries that enhance Hadoop's capabilities and ease of use, like [32]. These contributions from various entities have significantly enriched the Hadoop ecosystem, fostering innovations that benefit a wide array of users and applications across industries over the last decade.

\subsection{Storage}
The Hadoop ecosystem offers a variety of storage options tailored to different needs and data types, providing a flexible and robust environment for handling big data. At the core is the Hadoop Distributed File System (HDFS), designed for reliable and scalable storage of large datasets across clusters of commodity hardware. Beyond HDFS, Hadoop integrates with other storage systems to cater to diverse requirements. Cloud storage options like Amazon S3 or Google Cloud Storage can be seamlessly integrated with Hadoop, providing flexibility and scalability while optimizing costs.

HDFS is the main choice for storage in Hadoop ecosystems. It is a distributed file system that provides a solution to the problem of storing data across multiple machines in a cluster. Its design has a master/slave architecture with a single Name Node as the master server which manages the file system namespace and regulates access to files by clients. The slaves are a number of Data Nodes, usually one per node in the cluster, which manage storage attached to the nodes that they run on [28]. The system is designed to ensure fault tolerance, scalability and efficient storage and retrieval of large-scale datasets.

In HDFS, data is stored in the form of blocks or chunks, which are typically large and set to a default size of 128 MB in Hadoop 2.x and 3.x. By default, each block is replicated three times, though this configuration can be customized on a per-file basis. These replicas are then distributed across nodes in the Hadoop cluster, as shown in figure 2, which ensures both fault tolerance and efficient processing of data. The HDFS NameNode keeps track of the location of these chunks and is responsible for managing file system namespace operations. When a client application requests location information from the Namenode, it responds with the relevant chunk handle and chunk locations. If a certain location is unavailable, the client automatically selects the next available location and retries the request. This behavior is critical in maintaining fault tolerance in HDFS.

\begin{figure}[ht]
\centering
\includegraphics[width=.70\textwidth]
{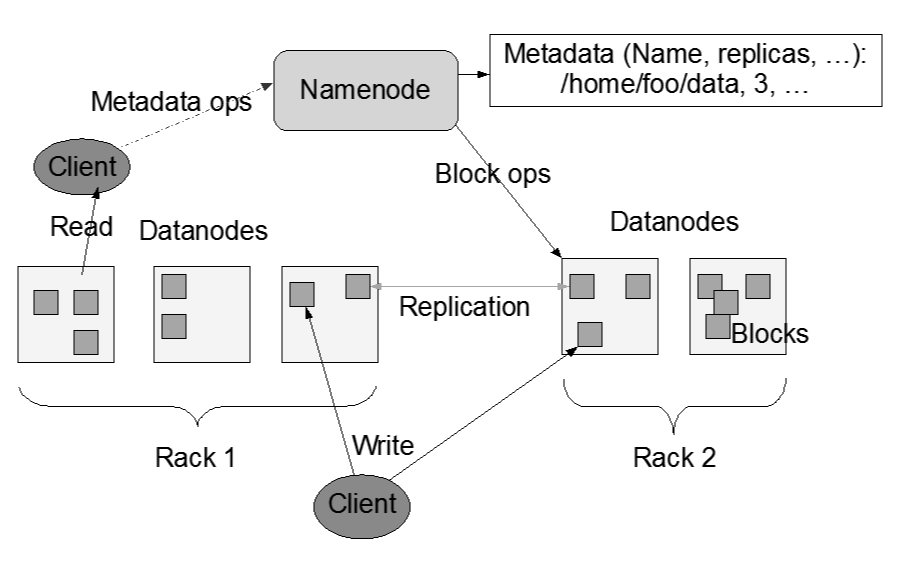}
\caption{HDFS architecture}
[23]
\end{figure}

Further enhancing this replication scheme is HDFS's rack-awareness feature, an intelligent data placement strategy that extends beyond mere node-level redundancy. Rack-awareness considers the physical configuration of the cluster, ensuring that data replicas are not only distributed across different nodes but also across different racks. This spatial dispersion of data helps mitigate the risk of data unavailability or loss during rack-level failures, enhancing the overall resilience of the system. Moreover, the rack-awareness capability can be extended to encompass multiple data centers, enabling HDFS to support a geo-distributed architecture [23]. By distributing replicas across different data centers, HDFS can provide a robust disaster recovery solution, enabling HDFS to continue operations even if an entire data center goes offline and significantly reducing the dependency on traditional backup mechanisms.

There are alternatives to HDFS. Alluxio, known as a virtual distributed storage system, acts as a layer between computation frameworks and storage systems, enabling faster data access by abstracting the underlying storage and providing data in-memory [33], which is particularly beneficial in big data and machine learning scenarios. Lustre, widely adopted in high-performance computing (HPC), is a scalable file system that supports large-scale, high-bandwidth storage environments, making it suitable for intensive data processing tasks [49]. Ceph offers a highly reliable and scalable storage solution, supporting object, block and file storage under one whole system [2], ideal for cloud storage and big data applications due to its fault tolerance and self-healing capabilities.

XtreemFS is a distributed file system designed to support fault tolerance, replication and scalability, allowing for flexible data access across geographically distributed networks, making it well-suited for grid and cloud-based applications [26]. Lastly, GlusterFS stands out for its ease of use and configurability, offering a scalable network filesystem that excellently serves applications requiring high-capacity storage and high-speed access to data [13].

\subsection{Resource Manager}
In the Hadoop ecosystem, resource management is crucial for orchestrating and optimizing the utilization of computational resources across clusters. YARN (Yet Another Resource Negotiator) stands out as the primary resource manager within Hadoop, designed to efficiently allocate resources for various applications, thereby enhancing the system's overall performance and scalability. It separates the job scheduling and resource management functions from the data processing component, allowing for a more versatile and robust processing environment. Yarn achieves this by introducing a master daemon known as the Resource Manager, which handles the allocation of computational resources among all the applications in the system, and a per-application Application Master that manages the application's lifecycle and resource needs within the framework [62].

Besides YARN, Apache Mesos is another notable resource manager, known for its fine-grained sharing capabilities and ability to run Hadoop alongside other applications. As stated by [39], a framework running on top of Mesos consists of two components: a scheduler that registers with the master to be offered resources and an executor process that is launched on agent nodes to run the framework's tasks. While the master determines how many resources are offered to each framework, the frameworks' schedulers select which of the offered resources to use. When a framework accepts offered resources, it passes to Mesos a description of the tasks it wants to run on them. In turn, Mesos launches the tasks on the corresponding agents. According to [24], a framework can reject resources that do not satisfy its constraints in order to wait for ones that do. Thus, the rejection mechanism enables frameworks to support arbitrarily complex resource constraints while keeping Mesos simple and scalable.

While not traditionally part of Hadoop, Docker Swarm, and specially Kubernetes, have emerged as powerful platforms for container orchestration, capable of managing Hadoop containers and enabling more dynamic resource allocation and scaling.

\subsection{Distributed Processing}
MapReduce is a programming model and software framework that enables the distributed processing of large data sets on clusters of computers. As the large datasets are divided into smaller pieces in HDFS, tasks can then be processed in parallel on the different Data Nodes.

MapReduce consists of two primary operations: map and reduce. The map operation takes a set of input data and produces an intermediate set of key-value pairs. The reduce operation then takes these intermediate key-value pairs and combines them to produce a final set of output values. The map and reduce operations are both designed to be highly parallelizable, which makes MapReduce well-suited for distributed computing environments.

\begin{figure}[ht]
\centering
\includegraphics[width=.60\textwidth]{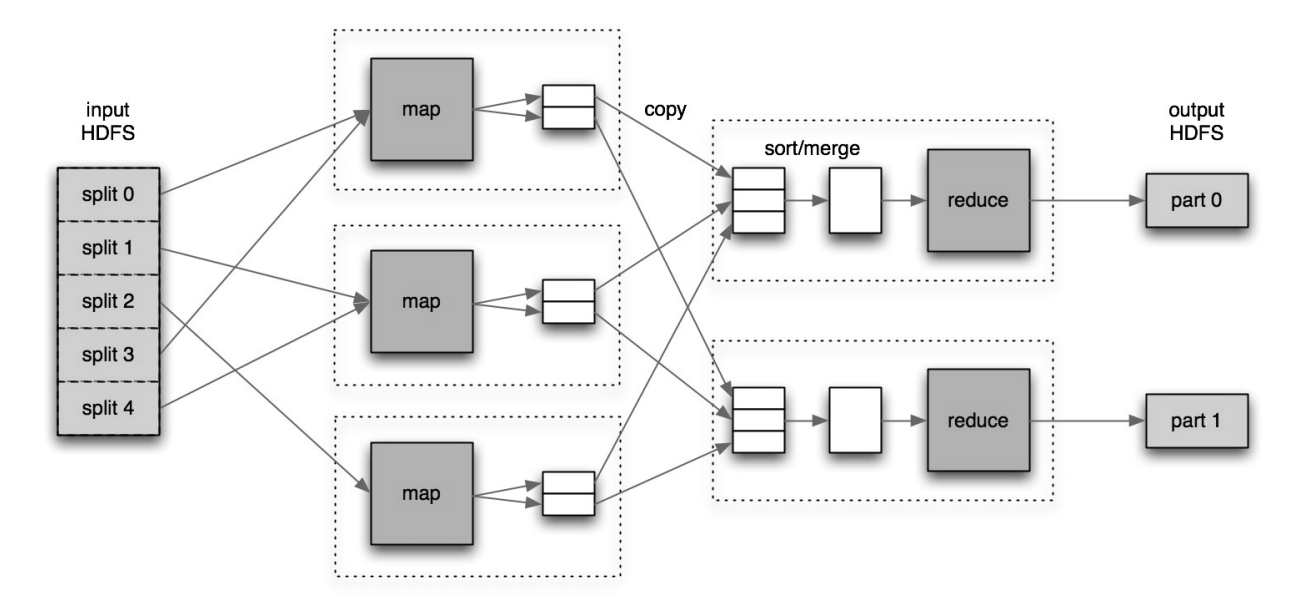}
\caption{Distributed Processing with MapReduce} [60]
\end{figure}

As illustrated in figure 3, the input comprises three lines which are partitioned into three parts during the Splitting phase and forwarded to individual Map-type tasks. During the Mapping phase, each Map task operates on its respective input on a cluster node and generates a key-value structure. In the specific example, the Map task returns the word count. Subsequently, in the Shuffling phase, data is transferred from the Mappers nodes to the Reducers nodes. Prior to the transfer, a sort is performed on all the keys, which ensures that the same keys are transferred to the same Reducer. The Reduce phase aggregates the data. In the example, this aggregation is the summation of the values associated with each key, which results in the number of occurrences of each word in the original input. Finally, the outputs of each Reducer are combined to generate the ultimate result.

\subsubsection{Spark}
MapReduce jobs can be really complex and require many linked Map and Reduce tasks, so many files will have to be generated at runtime. As these files are generated on disk, processing becomes I/O intensive and the cost of complex operations becomes high enough that processing performance is degraded. Spark was created to solve this performance degradation problem in MapReduce, replacing disk structures with distributed memory structures, and eliminating the I/O cost associated with multiple disk reads and writes.

The memory structures were called Resilient Distributed Datasets in the seminal article [64], but today they are called just Dataframes. As we can see in Figure 4, the intermediary files generated by Map and Reduce tasks are placed in memory.

\begin{figure}[ht]
\centering
\includegraphics[width=.60\textwidth]{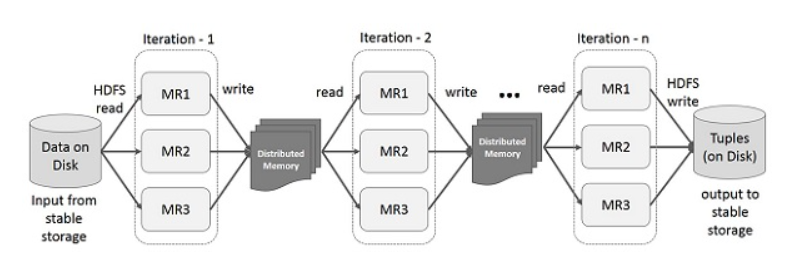}
\caption{Apache Spark Resilient Distributed Datasets} [55]
\end{figure}

Therefore, the conventional method of updating files on disk has been replaced in Spark with in-memory updates distributed across the nodes of the cluster. This feature enables the Map and Reduce tasks to directly access the intermediate processing data in the memory of the cluster nodes, leading to a significant improvement in the performance of the previously inefficient MapReduce jobs. Besides the superior utilization of memory for processing operations instead of disk, one of the distinctive attributes of Spark is its capability to execute interactive applications, such as SQL queries. Spark is considered an extension of MapReduce, which is confined to batch applications, and it provides support for interactive applications including Machine Learning, Streaming, SQL queries and Graph processing.

\subsection{Distributed Programming}
Distributed programming in the Hadoop ecosystem is a fundamental concept that enables the processing of large data sets across clusters of computers using simple programming models. It is designed to scale up from a single server to thousands of machines, each offering local computation and storage. Key components like Hadoop MapReduce and Apache Spark facilitate distributed programming by abstracting the complexity of underlying network communication, data distribution and fault tolerance. Both frameworks support writing applications in various languages, such as Java, Python and Scala, enabling developers to leverage distributed programming to process vast amounts of data efficiently. 

This capability is crucial in the Hadoop ecosystem, allowing it to serve as a backbone for big data processing and analytics, enabling batch-based parallelized work to be performed on a cluster of multiple nodes [25], enabling insights and decisions based on massive datasets that would be infeasible to process on a single machine. Java, being the language Hadoop itself is written in, offers the most direct access to Hadoop's APIs and is widely used for MapReduce programming. It provides strong integration, performance, and control over the Hadoop ecosystem components, making it a preferred choice for developers seeking deep customization and optimization in their Hadoop applications.

Python is another popular choice in the Hadoop ecosystem, largely due to its simplicity and the extensive availability of data processing and machine learning libraries. Tools like PySpark, the Python API for Spark, and Hadoop Streaming, which allows writing MapReduce programs in any language that can read from stdin and write to stdout (like Python), have made Python a powerful and flexible language for distributed data processing in Hadoop. Python's readability and ease of use make it an attractive option for data scientists and analysts who may not have a deep programming background but need to work with large data sets. Scala, on the other hand, has gained popularity in the Hadoop ecosystem primarily due to Apache Spark, which is written in Scala. This language offers concise syntax and functional programming capabilities, making it ideal for data processing tasks. Scala's seamless integration with Spark allows developers to write more readable and concise code, which is particularly beneficial for complex data transformation and analysis tasks. Additionally, Scala's interoperability with Java means that developers can easily integrate Scala-based Spark applications with existing Java-based Hadoop components.

\subsection{SQL-based engines}
Hive, Impala and Presto are all SQL-based engines that allow for querying data stored in distributed systems, but they differ significantly in their architecture and performance characteristics. While they can interact with SQL metastores (like the Hive metastore), they are not SQL metastores themselves. Instead, they are query engines that can utilize metadata from a metastore to execute queries.

Hive allows users to execute SQL-like commands, known as HiveQL, to analyze data stored in Hadoop's HDFS or other compatible storage systems. Hive translates SQL queries into MapReduce jobs, which can be slower due to the overhead of MapReduce. However, it also supports execution engines like Tez and Spark for faster processing. It is typically used for batch processing tasks and data warehousing applications where high latency is acceptable. One key aspect of Hive is that it provides two types of table structures: external and internal (managed) tables, each serving distinct purposes and use cases within data processing workflows. Internal tables are managed entirely by Hive, meaning that Hive controls the data lifecycle, including storage and metadata. When an internal table is deleted, Hive deletes both the table schema and the data itself. This tight coupling between Hive and the data makes internal tables a suitable choice when full data management is needed within the Hive ecosystem. 

\begin{figure}[ht]
\centering
\includegraphics[width=.60\textwidth]
{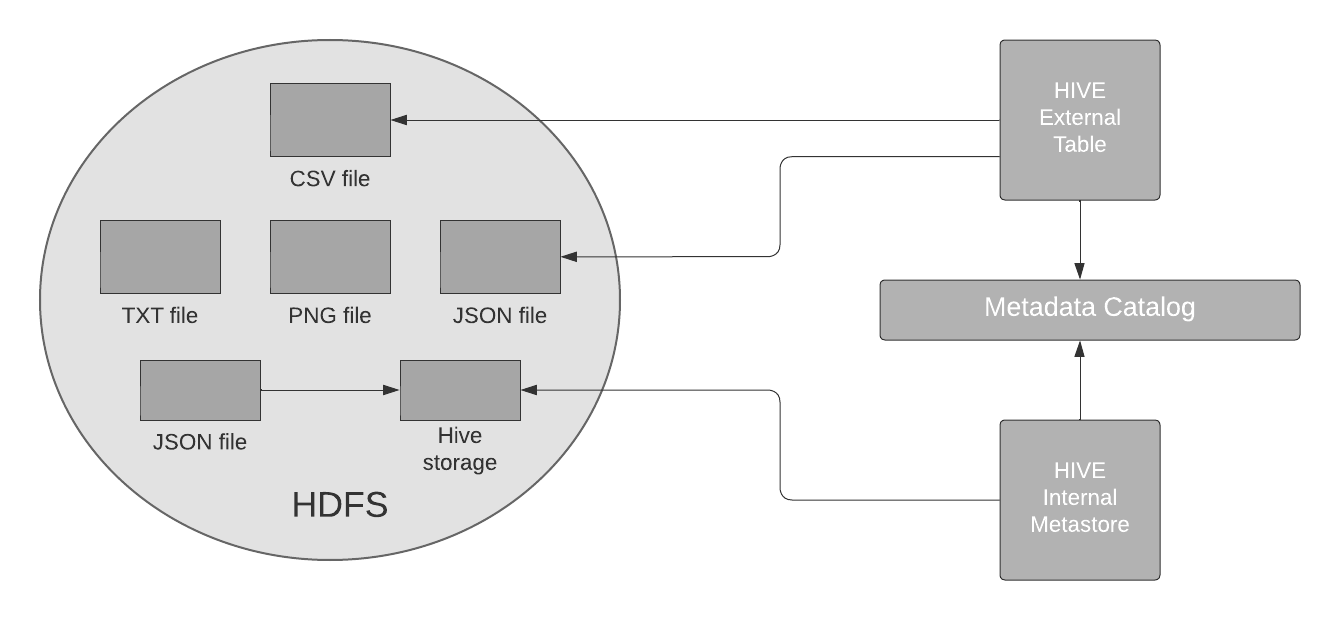}
\caption{Apache Hive internal and external tables} 
\end{figure}

On the other hand, as shown in figure 5, external tables allow Hive to access data that is stored outside of Hive's purview, typically in a location specified by the user in the HDFS or other file system. When an external table is deleted, Hive only removes the schema and leaves the data untouched [53]. This feature is particularly useful when the data needs to be shared across different applications or when there is a need to maintain the data independently of Hive's management lifecycle. For example, a Hive external table can be created to map to a CSV file stored in HDFS, allowing users to execute a SELECT statement to query the data as if it were in a traditional table, while the same CSV file can concurrently be accessed by a MapReduce job for other data processing tasks.

Impala and Presto are designed for interactive SQL queries. Impala was developed by Cloudera and is a massively parallel processing (MPP) SQL query engine that bypasses MapReduce, directly accessing the data stored in HDFS, HBase or Amazon S3, which allows for faster query execution compared to Hive. Presto was developed by Facebook and operates in-memory, without relying on MapReduce or HDFS. It can query data from multiple sources, including Hive, HBase, relational databases and even proprietary data stores [52]. Both Impala and Presto can integrate with the Hive metastore for metadata, but is not limited to Hadoop-based data sources.

As part of the efforts to improve Apache Hive, the feature LLAP (Live Long and Process) became available as part of Hive 2.0, which was released in early 2016. It is an addition to Apache Hive that significantly enhances its capabilities for interactive SQL queries, making it more competitive with tools like Impala and Presto, which are designed for low-latency SQL querying on Hadoop data. LLAP introduces a daemon service that enables in-memory caching, which helps to avoid reading data from disk on repeated queries [34]. This persistent service holds data in memory and processes across queries, which drastically reduces query latency by eliminating the need to read data from disk for every query.

\subsection{NoSQL data stores}
In the Hadoop ecosystem, NoSQL databases, often referred to as NoSQL data stores, play a crucial role in managing and processing vast amounts of unstructured or semi-structured data. These databases are designed to overcome the limitations of traditional relational databases, particularly in terms of scalability, flexibility and performance when dealing with big data. HBase, Giraph and Cassandra are prominent examples within this ecosystem.

Apache HBase is a distributed, scalable, big data store built on top of the Hadoop Distributed File System (HDFS). It is a column-oriented NoSQL database that is designed for quick read/write access to large datasets. HBase shines in scenarios where real-time read/write access and high throughput are required, making it a popular choice for big data applications that need to handle massive amounts of data across thousands of nodes [21]. It supports row-keyed data, which allows for easy retrieval of rows by key, and it's often used for storing sparse data sets where many columns are empty. Besides, there's an integration between Hive and HBase, which supports originally HiveQL statements to access HBase tables for both read and write. It is even possible to combine access to HBase tables with native Hive tables via joins and unions [22].

Although not exclusively part of the Hadoop ecosystem, Apache Cassandra is a highly scalable NoSQL database that can integrate with Hadoop for big data processing. Cassandra provides robust support for clusters spanning multiple datacenters, with asynchronous masterless replication allowing low latency operations for all clients. Its data model offers the convenience of column indexes with the performance of log-structured updates, strong support for denormalization and powerful built-in caching. According to [9], in relational databases, data is placed in normalized tables with foreign keys used to reference related data in other tables. Queries that the application will make are driven by the structure of the tables and related data are queried as table joins. In Cassandra, data modeling is query-driven. The data access patterns and application queries determine the structure and organization of data which then used to design the database tables.

Neo4j, a prominent graph database, is designed to handle highly connected data and complex queries with agility. Unlike traditional databases, which may struggle with deep join operations, Neo4j excels in exploring relationships within data, making it ideal for use cases like social networks, recommendation engines and fraud detection. Its graph model allows for data to be stored in nodes and relationships, facilitating efficient traversal and querying of connected structures. Neo4j's Cypher query language provides an intuitive way to express graph patterns and retrieve data, enhancing the database's usability and performance in graph-related computations. According to [42], Neo4j is engineered to capitalize on the advantages of graph processing, offering significant improvements in speed and flexibility for applications that require the analysis of interconnected data. This capability underscores Neo4j's suitability for scenarios where relational databases might not provide optimal performance or where the data's relational aspects are critical to the application's core functionality.

Within the Hadoop ecosystem, a variety of NoSQL databases offer specialized functionalities that enhance big data processing capabilities. MongoDB and Couchbase, both document-oriented databases, integrate seamlessly with Hadoop, enabling data synchronization and analytics with their flexible schema and efficient querying. Accumulo, built on Hadoop, provides a secure key value store with fine-grained access control, ideal for sensitive data handling. Similarly, Redis, primarily an in-memory data store, complements Hadoop's processing power by caching results for faster access, whereas Riak offers distributed storage solutions with high availability and fault tolerance.

On the other hand, Aerospike, known for its high performance, serves as a key-value store that supports real-time data processing, making it suitable for applications requiring rapid data access and analytics. These NoSQL data stores, when integrated with Hadoop, provide robust solutions for managing and analyzing diverse data sets, ranging from structured to unstructured data. They cater to various industry needs, from real-time bidding systems to secure data management, demonstrating the flexibility and scalability essential for contemporary big data ecosystems.

Finally, Elasticsearch, often recognized for its powerful search and analytics capabilities, also functions effectively as a NoSQL database. It manages and stores data in a schema-less fashion using JSON documents, a characteristic that aligns with the NoSQL paradigm. This feature enables Elasticsearch to handle vast volumes of unstructured and semi-structured data efficiently, providing scalability and flexibility in data indexing and search operations. Its distributed nature allows for high availability and resilience, facilitating rapid data retrieval and real-time analytics across large datasets. According to [58], Elasticsearch's capability to perform complex searches and analytics at scale, coupled with its document-oriented storage mechanism, exemplifies its role as a NoSQL database, addressing use cases that traditional relational databases may find challenging. This dual functionality as both a search engine and a NoSQL data store makes Elasticsearch a versatile tool in the realm of big data and real-time analytics applications.

\section{Polyglot Persistence}
Polyglot persistence refers to the practice of using different data storage technologies to handle varied data storage needs within the same application. Instead of trying to fit all data into a single storage model, developers choose the best storage solution for each workload or application, such as using a document store for JSON data, a graph database for interconnected data, and a relational database for a transactional system. According to [29], polyglot persistence refers to the use of different types of databases in one application system, instead of forcing all data to fit into only one database type.

For example, [31] shows that an e-commerce solution may use a document store to manage orders data due to its flexible schema for semi-structured data. Simultaneously, a graph database could be employed to analyze and recommend products based on user relationships and interactions, while a key-value store for session management and shopping cart data. Additionally, a relational database could be used for transaction processing, where ACID properties are crucial, like money balance data. This setup ensures that each aspect of the e-commerce platform leverages the most suitable data storage and retrieval mechanisms, as illustrated in Figure 6.

\begin{figure}[ht]
\centering
\includegraphics[width=.70\textwidth]
{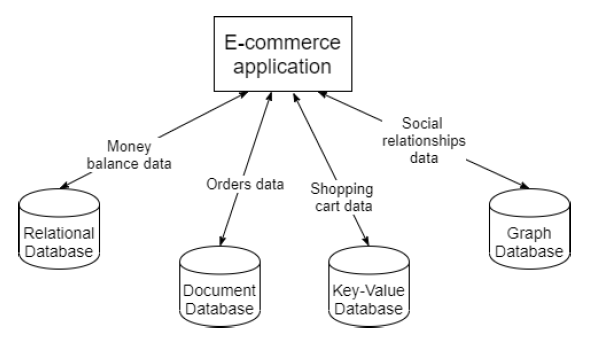}
\caption{Polyglot persistence in E-commerce application} [31]
\end{figure}

While polyglot persistence offers flexibility, it typically requires the application layer to manage the integration and interaction between these diverse data stores. One explicit option is for the application itself to orchestrate these interactions, meaning the application code directly handles the logic for when and how to interact with each data store, as well as how to integrate and process the data from different sources. This approach provides granular control but can increase the complexity of the application code and tightly couple it with the data layer.

On the other hand, using mediator systems or middleware can abstract away the complexities of direct interaction between the application and the data stores. Middleware solutions can provide a unified interface or communication layer that sits between the application and the databases, handling the necessary translations and data transformations. [48] proposes a Polyglot Persistence Mediator (PPM), which allows for runtime decisions on routing data to different backends according to schema-based annotations. As the authors state, PPM acts as a broker between applications and backend databases. Applications use a defined interface to issue queries, CRUD operations, transactions, and other operations to the mediator in a database-agnostic fashion. Based on the routing model, the mediator selects the appropriate database and transforms the incoming operation to database-specific operations.

Other implementations of mediators include data virtualization platforms and data management frameworks, providing an abstraction layer over heterogeneous data sources and allowing users to access and query data without needing to know where and how the data is stored. Data virtualization tools often provide capabilities to federate queries, cache results for performance, and translate queries between different query languages. For example, [8] offers a set of tools that include a SQL parser, an optimizer, and a query execution engine that can interpret, optimize, and execute queries across a wide array of underlying data sources, such as Cassandra, MongoDB, and Redis. According to [20], Calcite represents its queries via a relational operator tree, implementing the base SQL language with a subset of extensions and supporting heterogeneous data models.

\subsection{Polyglot Data Stores}
Polyglot Data Stores have emerged as a sophisticated approach to data management, aimed at utilizing the most effective data storage technology for different data types and use cases within the same application. The term \textit{polyglot persistence}, coined by Martin Fowler [38], encapsulates the notion of applying multiple data storage technologies as dictated by the specific needs of each data model within an application. According to [46], the polyglot approach leverages the strengths of various storage technologies, such as the document model's flexibility, the wide column store's scalability, the key-value store's speed, and the graph database's connected data efficiency. By strategically employing these diverse technologies, polyglot data stores enable applications to meet complex and varied data persistence demands more effectively than a one-size-fits-all solution.

Often called Polystores, they can be categorized based on their architecture, the supported data models, the nature of their data integration, and the level of transparency they offer to end-users in handling heterogeneous data sources. Based on their architecture, polystores can be integrated or federated. In the integrated approach, different data models are integrated into a single system that manages the data in a unified manner. This approach aims to provide a seamless experience across different data models. The federated approach involves a collection of disparate data management systems, each supporting different data models, but interconnected through a federated layer that allows for integrated query processing and data retrieval.

Based on the supported data models, a multimodel data store supports multiple data models but might not offer full flexibility or optimization for each model. On the other hand, hybrid polystores are optimized to support two or more specific data models, offering specialized functionalities and optimizations for each supported model.

[20] indicates four polystores categories: Federated Systems, Polylingual Systems, Multistore Systems, and Polystore Systems. Federated and Polylingual Systems use a homogeneous set of data stores beneath their mediation layer, whereas Multistore and Polystore systems rely on heterogeneous ones. Furthermore, Federated and Multistore Systems only offer one query interface, while Polylingual and Polystore Systems provide many interfaces. It's important to note that these categories are not universally agreed upon in the literature, and many different classifications exist. However, all approaches attempt to distinguish between homogeneous and heterogeneous data sources and the tools available for data retrieval.

Federated Systems, such as Apache Calcite [8], allow querying across multiple databases without moving data, treating them as if they were a single entity. Despite the underlying databases being homogeneous, the federated system abstracts their complexities to offer a unified query interface. On the other hand, Polylingual Systems support multiple query languages or data models within the same system, catering to a homogeneous set of data stores. An example could be a database that supports both SQL for relational data and CQL for columnar data, like ScyllaDB [50].

Multistore Systems involve heterogeneous data stores and provide a single query interface to access them. An example is Presto, which can query data from HDFS, S3, Cassandra, MySQL, and many other sources through a single SQL-like interface. Polystore Systems integrate heterogeneous data sources and offer multiple query interfaces to suit the data models of the integrated stores. An example is Polybase, which allows SQL Server to execute queries that combine data from SQL Server and Hadoop or from other databases like Oracle or Teradata.

\subsection{CAP Theorem}
The relationship between Polyglot Persistence and the CAP Theorem becomes particularly salient in the context of distributed data stores, as the theorem provides a crucial framework for understanding the inherent trade-offs in such systems. In a polyglot persistence setup leveraging distributed data stores, each system may prioritize different aspects of the CAP theorem - Consistency, Availability or Partition Tolerance-based on its design principles and operational characteristics. This prioritization is pivotal when selecting and integrating multiple data storage technologies to cater to specific use cases or data types within a distributed environment. Understanding these trade-offs is essential for architects to design a polyglot architecture that aligns with the application's requirements, ensuring that data is managed, accessed and synchronized effectively across distributed stores. By carefully choosing data stores that align with the desired CAP properties, architects can construct a resilient and adaptable polyglot persistence framework, enabling the application to efficiently handle diverse data demands and enhance overall system performance in a distributed setting. 

According to [29], The CAP (Consistency, Availability, Partition Tolerant) conjecture explains why SQL is not sufficient for a distributed system. Since the ACID characteristics are difficult to fulfill in a distributed manner, many researchers are considering new types of database systems with relaxed characteristics for implementation in distributed systems. Indeed, as [20] states, the architecture of polyglot systems can be classified according to the CAP theorem. Since they can choose and set up their underlying data stores according to the user's current needs, they can theoretically be CA, CP and AP at the same time - never all at once for the same application.

\section{Polyglot nature of the Hadoop ecosystem}
While not polystores in the strictest sense, data lakes can function in a polystore-like manner by storing raw data in various formats and enabling diverse analytical and processing tools to operate on this data. In such platforms, the data lake serves as a centralized repository, from which data can be directed to the most suitable data store. Various engines and tools can then query and process the data as needed. This setup exemplifies both polyglot persistence, by accommodating raw data in numerous formats, and polyglot processing, by allowing a diverse array of analytical and processing tools to interact with the data. 

The Hadoop ecosystem is increasingly utilized as a foundational technology for data lakes, especially within public cloud environments, where its ability to handle vast amounts of diverse data shines. Major cloud providers, such as Amazon Web Services (AWS), Microsoft Azure and Google Cloud Platform (GCP), offer Hadoop-based services that simplify setting up data lakes. Amazon Web Services (AWS) offers Amazon Elastic MapReduce (EMR), a cloud-native big data platform that allows processing vast amounts of data quickly and cost-effectively across resizable Hadoop clusters. Microsoft Azure provides Azure HDInsight, a fully managed cloud service that makes it easy, fast and cost-effective to process massive amounts of data. It supports a broad range of Hadoop ecosystem components, including Apache Spark, HBase and Kafka. Google Cloud Platform (GCP) offers Cloud Dataproc, a fast, easy-to-use, fully managed cloud service for running Apache Spark and Apache Hadoop clusters. Oracle offers Big Data Service, which is a Hadoop-based service designed to provide a comprehensive and secure big data platform. Users can manage their Hadoop and Spark workloads with scalable clusters, integrated with OCI's storage and database services. IBM's Analytics Engine is a service that combines Apache Hadoop and Apache Spark to offer a powerful environment for analyzing and processing vast datasets. It simplifies the process of managing hardware and software resources, allowing users to focus on data analysis and application development. And Cloudera, one of the original Hadoop distribution companies, offers its services on various cloud platforms. Cloudera Data Platform (CDP) can be deployed on AWS, Azure and GCP, providing a unified, secure and open data platform that covers the entire data lifecycle.

These cloud-based data lakes leverage Hadoop's robust ecosystem, including storage options like HDFS or cloud object storage (S3 on AWS, Blob Storage on Azure and Cloud Storage on GCP) and processing frameworks like MapReduce and Spark. The elasticity of the cloud allows organizations to scale their data lakes up or down based on demand, optimizing costs and performance. Additionally, cloud providers integrate Hadoop with various analytics and machine learning services, enhancing the data lake's capabilities and making it a central hub for data ingestion, storage, processing and analysis. This seamless integration in the cloud enables businesses to harness the power of big data analytics more efficiently, driving insights and innovation while leveraging Hadoop's proven scalability and flexibility.

Figure 7 shows a diagram related to a Polyglot Big Data processing architecture based on Hadoop. Starting with the Ingestion Phase, data enters the system and is first deposited in the Landing Zone. This is a preliminary staging area where raw data is collected before any processing. It serves as an initial checkpoint where data can be quickly captured and stored in its native format. From the Landing Zone, if necessary, data undergoes a transformation process to become more structured and usable; it is then moved to the Enriched Zone. This zone represents an intermediate layer where data has been cleansed, enriched and possibly reformatted to ensure higher quality and readiness for analysis. This transformation may be crucial for preparing the data for efficient querying and processing. 

\begin{figure}[ht]
\centering
\includegraphics[width=.70\textwidth]
{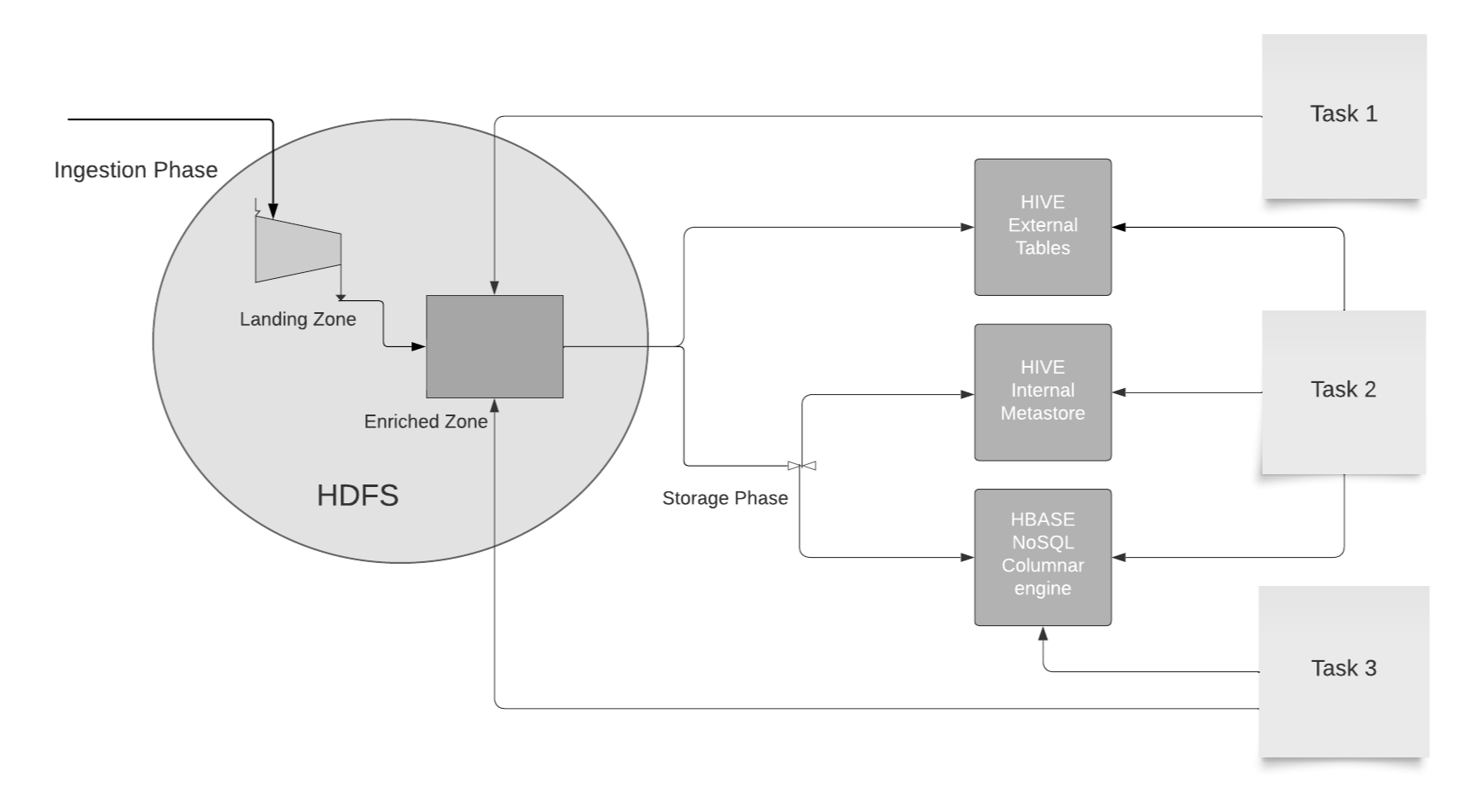}
\caption{Polyglot Architecture}
\end{figure}

In the Storage Phase, data is routed to distinct storage systems based on its characteristics and the intended use. In this instance, the storage options utilized are Hive external tables, Hive internal tables and HBase tables. Progressing to the final stage of the workflow, Tasks 1, 2 and 3 each represent specific analytical or processing jobs within the platform. The data store chosen for each task is based on its compatibility with the task's requirements. Task 1 might interact directly with an enriched JSON file stored on HDFS, ideal for analytical and querying tasks that do not alter the raw data. Task 2 could necessitate access to multiple data stores, combining data retrieved from these varied sources. Meanwhile, Task 3 is likely to capitalize on HBase, particularly for real-time processing and applications that demand rapid data access, merging data sourced straight from HDFS.

Polyglot Data Processing in this context involves using various data stores and processing technologies suited to different data types, from structured data in relational databases to unstructured data in NoSQL systems. It combines multiple programming languages and tools for diverse data management, all centralized around HDFS. This approach allows for efficient and scalable data processing, especially in complex areas like healthcare and social network analysis. Moreover, it provides the flexibility to use specific storage solutions like HBase for certain tasks, thereby optimizing the data processing pipeline for both depth of analysis and efficiency.

\section{Use Cases}
In this section, we provide real-world examples or case studies that illustrate the effective use of the polyglot approach using the Hadoop ecosystem, showcasing how different tools may be combined to solve specific problems in big data processing.

\subsection{Healthcare}
Consider a healthcare analytics platform designed to provide comprehensive insights into patient care, treatment outcomes and operational efficiency for a network of hospitals, as shown in figure 8. This platform could greatly benefit from using several storage and processing techniques within the Hadoop ecosystem to handle various data types and analysis needs.

To ingest real-time data from IoT devices used in patient monitoring, such as heart rate monitors and blood glucose sensors, Kafka can handle the high-throughput data streams these devices generate. The data can then be processed using Apache Storm or Spark Streaming to detect anomalies or critical conditions in real-time, triggering alerts for immediate medical intervention. Use HBase for storing patient records and clinical data. HBase's columnar storage model is well-suited for this kind of data, allowing for efficient read/write operations and easy scalability. It can store vast amounts of structured and semi-structured data, making it ideal for handling electronic health records (EHRs), lab results and imaging data.

To create a data warehouse that consolidates various data sources, providing a unified view for more complex analytics, Hive allows for querying this data using SQL-like queries, making it accessible for data analysts and decision-makers to generate reports, conduct trend analysis and support strategic planning. Additionally, the solution may use MapReduce and Spark for batch processing and analysis of historical data. This could involve analyzing treatment outcomes over time, patient recovery rates, or the effectiveness of different medications. The results can inform future treatment plans and healthcare policies.

\begin{figure}[ht]
\centering
\includegraphics[width=.70\textwidth]
{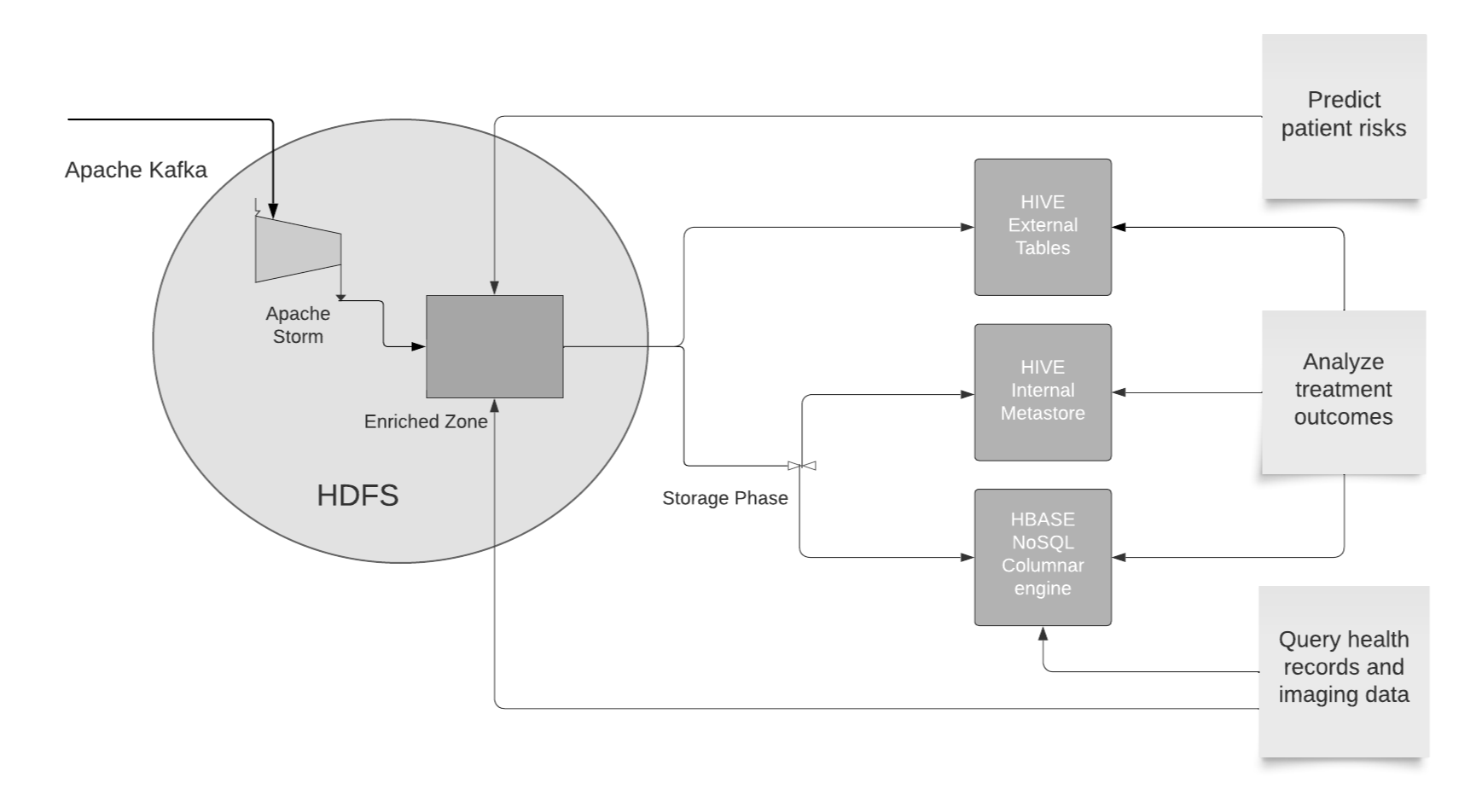}
\caption{Healthcare Polyglot Data Processing}
\end{figure}

As a complement, the solution may leverage Spark's MLlib for machine learning tasks, such as predicting patient readmission risks or identifying potential outbreaks of hospital-acquired infections. By training models on historical data, the platform can provide predictive insights that help improve patient care and operational efficiency. By integrating these diverse storage and processing technologies, the healthcare analytics platform can provide real-time monitoring, in-depth analysis and predictive insights, enhancing patient care and operational decision-making across the hospital network.

\subsection{Stock Market}
In order to process real-time stock market data feeds, one possible solution would leverage Apache Storm, utilizing its capabilities for fast, distributed and fault-tolerant stream processing to analyze market dynamics like ticks, trades and quotes instantaneously. This real-time analysis enables the platform to deliver prompt insights and alerts reflective of the current market conditions. Integrate Apache Flume to gather, aggregate and transport substantial volumes of log data, such as trade executions and order books, into HDFS, ensuring reliable and distributed data capture for subsequent processing, as shown in figure 9.

To enhance the platform's data handling and integration capabilities, we may use Apache NiFi, which facilitates data flow automation and management, allowing for the enrichment of real-time and historical market data by assimilating external data sources-news feeds, economic indicators or social media insights that could impact stock valuations. For processing and in-depth analysis of historical stock market data, employ Apache Spark to leverage its in-memory processing for swift computations on data stored in HDFS, aiding in trend analysis, volatility studies and the back-testing of trading strategies.

In the storage layer, we use Apache Cassandra to manage the influx of real-time data and HBase to manage the historical data, providing scalable and efficient storage solutions that support the platform's analytical demands. for dynamic data exploration and visualization, utilize interactive tools like Apache Zeppelin or Jupyter, integrated with Spark, enabling analysts and data scientists to script in notebooks, conduct exploratory analysis, prototype models and visually represent stock market trends or anomalies. We may integrate Elasticsearch as an additional polystore component to enhance the platform's search and analytics capabilities. Renowned for its powerful full-text search, near real-time analytics and scalability, Elasticsearch can serve as an efficient tool for indexing and querying vast amounts of structured and unstructured data generated in the stock market. By storing time-series data, such as stock prices and trade volumes, in Elasticsearch, the platform can quickly perform complex queries, aggregate data and analyze trends over time.

\begin{figure}[ht]
\centering
\includegraphics[width=.70\textwidth]
{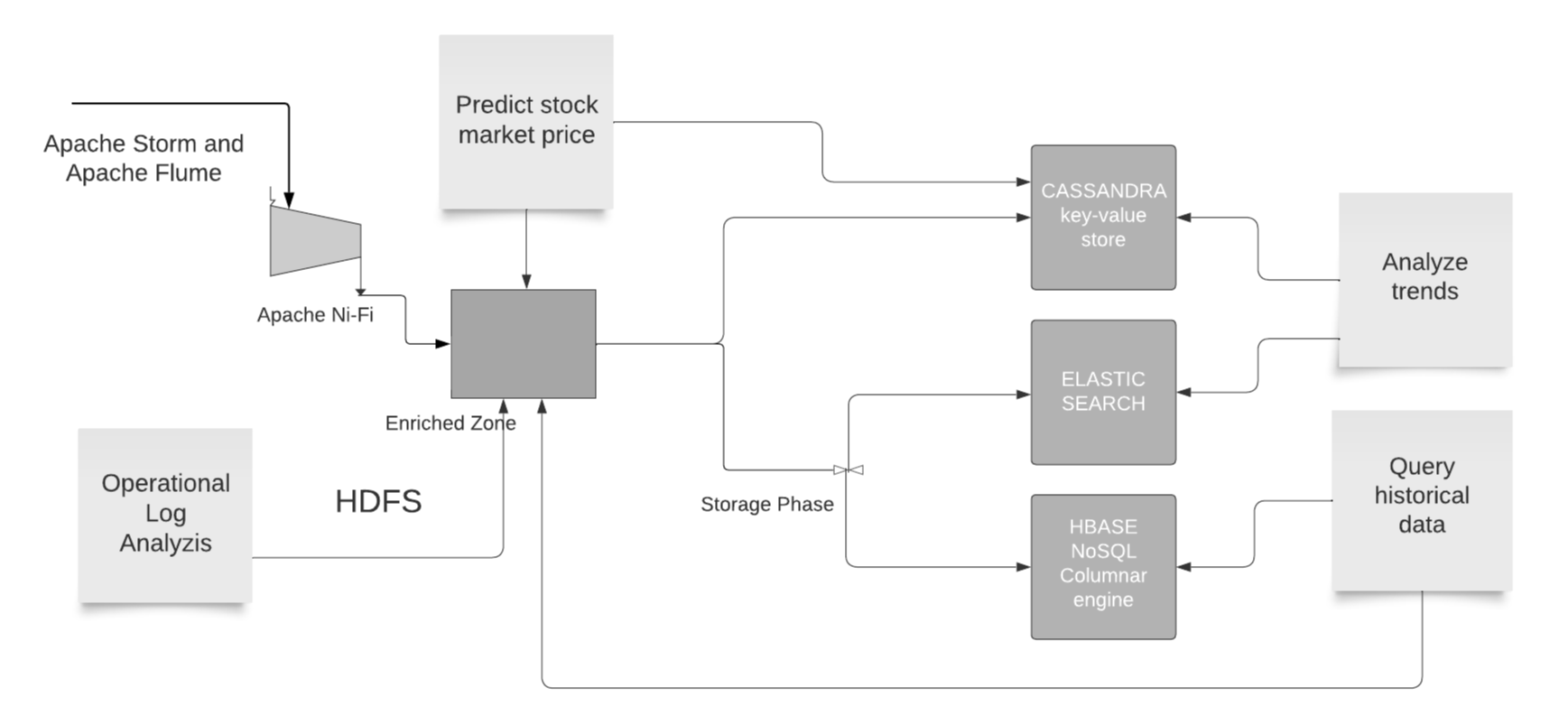}
\caption{Stock Market Polyglot Data Processing}
\end{figure}

Further, harness Spark's MLlib to develop and train machine learning models that predict stock price movements, discern market trends and evaluate risks by analyzing historical and real-time data, offering predictive insights to investors and traders for informed decision-making. Through these integrated technologies and strategic data storage solutions, the platform can achieve a comprehensive and nuanced understanding of market behaviors, facilitating advanced analytics for the financial sector.

\subsection{Social Networks}
Consider an application to analyse data that comes from a social network - Twitter, for example. to stream Twitter data in real-time using the Twitter API, we leverage Apache Kafka, which serves as the backbone for data ingestion, efficiently handling the high-volume and high-velocity data generated by Twitter users. It can capture tweets, retweets, likes and replies, ensuring a comprehensive dataset is available for analysis. We may use Apache Hive internal tables to store and manage this vast dataset in a structured format, as long as we may use Apache Hive external tables to consume data directly from HDFS, also in a structured view. Figure 10 shows a related diagram.

Hive's data warehousing capabilities allow the platform to organize Twitter data efficiently, enabling partitioning, indexing and querying of historical data. This organized data repository serves as the foundation for in-depth analytics and trend analysis over time. Implement Presto for fast, interactive querying of the Twitter data stored in Hive. Presto's distributed SQL query engine allows analysts and data scientists to perform exploratory data analysis, run ad-hoc queries and generate reports on various aspects of Twitter data, such as user engagement, hashtag popularity and sentiment trends, with low latency.

\begin{figure}[ht]
\centering
\includegraphics[width=.70\textwidth]
{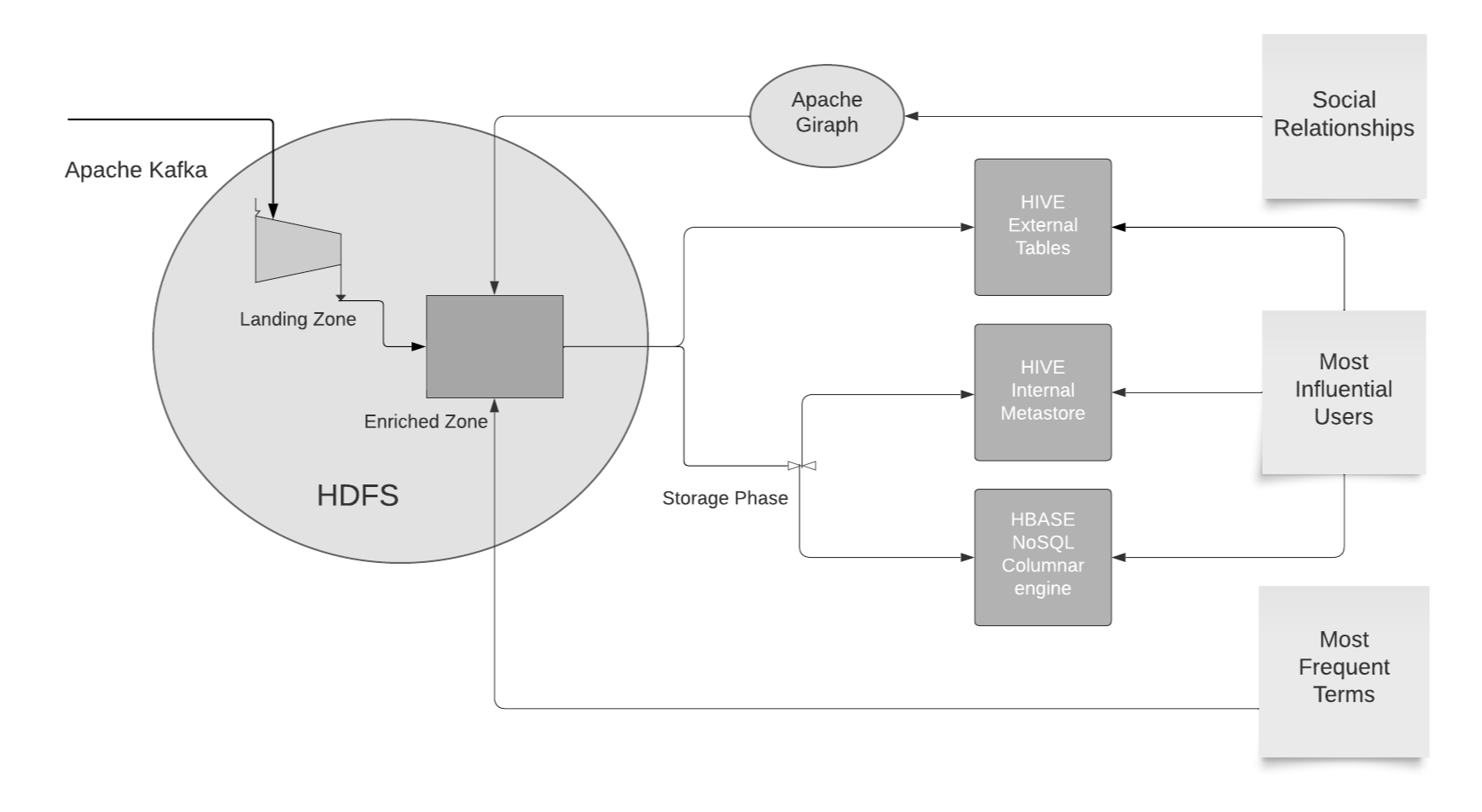}
\caption{Social Networks Polyglot Data Processing}
\end{figure}

One frequent job in this scenario is to count the most frequent terms found in tweets, a job can be implemented using the MapReduce framework to process large datasets efficiently. Initially, the MapReduce job would tokenize each tweet into words in the map phase and then count each occurrence in the reduce phase, providing a distributed mechanism to tally term frequencies across a vast collection of tweets. To enhance performance, this process can be further optimized by integrating Apache Spark, specifically leveraging Resilient Distributed Datasets (RDDs). Spark's RDDs offer a fault-tolerant, parallelized way to handle data, allowing for in-memory processing and reducing the need for disk I/O compared to traditional MapReduce. By utilizing Spark's RDD transformations and actions, the job can efficiently process and aggregate word counts in memory, significantly speeding up the analysis. This hybrid approach combines the robustness of MapReduce's distributed computing model with the speed and agility of Spark's in-memory processing capabilities, providing a powerful solution for real-time analytics on social media data.

Additionally, we integrate a graph processing framework, such as Apache Giraph or Neo4j, to model and analyze the Twitter network as a graph. In this graph, users can be represented as nodes and interactions such as follows, mentions and retweets as edges. This approach enables the platform to identify influential users, detect tightly-knit communities, understand the spread of information and uncover patterns in user interactions and content propagation. The platform can further employ machine learning models to perform sentiment analysis on tweet content, categorizing them into positive, negative or neutral sentiments. Combining sentiment data with community and interaction analysis, the platform can detect emerging trends, monitor brand perception and understand public reaction to events in real-time. The evaluation of the jobs related to this use case can be found in [1]. 

\subsection{Smart Cities}
Imagine a smart city initiative that aims to integrate and analyze data from various sources to enhance urban living, optimize city services and support decision-making processes. The initiative utilizes Apache Calcite as a polyglot persistence framework and Apache Spark for processing streaming data, addressing diverse data management and analysis needs.

Apache Spark is employed to process and analyze streaming data from various IoT devices and sensors deployed throughout the city. These devices continuously transmit data related to traffic conditions, public transportation, energy usage and environmental factors. Spark's ability to handle large-scale data streams in real-time is crucial for the smart city to monitor its infrastructure dynamically, detect anomalies and respond to events as they occur. For instance, Spark can analyze traffic flow data to optimize traffic light timings, reducing congestion during peak hours. It can also process data from environmental sensors to monitor air quality and trigger alerts if pollution levels rise above safe thresholds.

Apache Calcite acts as a unifying layer for various data storage systems within the smart city's infrastructure. It provides a flexible framework that allows the city's data engineers to query multiple data sources using a single SQL interface, regardless of the underlying data format or storage model. For example, Calcite can be used to integrate data from a relational store like Hive storing city taxes records, a NoSQL database like MongoDB with handling of large volumes of data for a smart waste management system and a time-series database like OpenTSDB holding real-time public transportation data. By offering a consistent query interface, Calcite simplifies data access and integration, enabling more efficient data analysis and management across the city's diverse data ecosystem.

\begin{figure}[ht]
\centering
\includegraphics[width=.70\textwidth]
{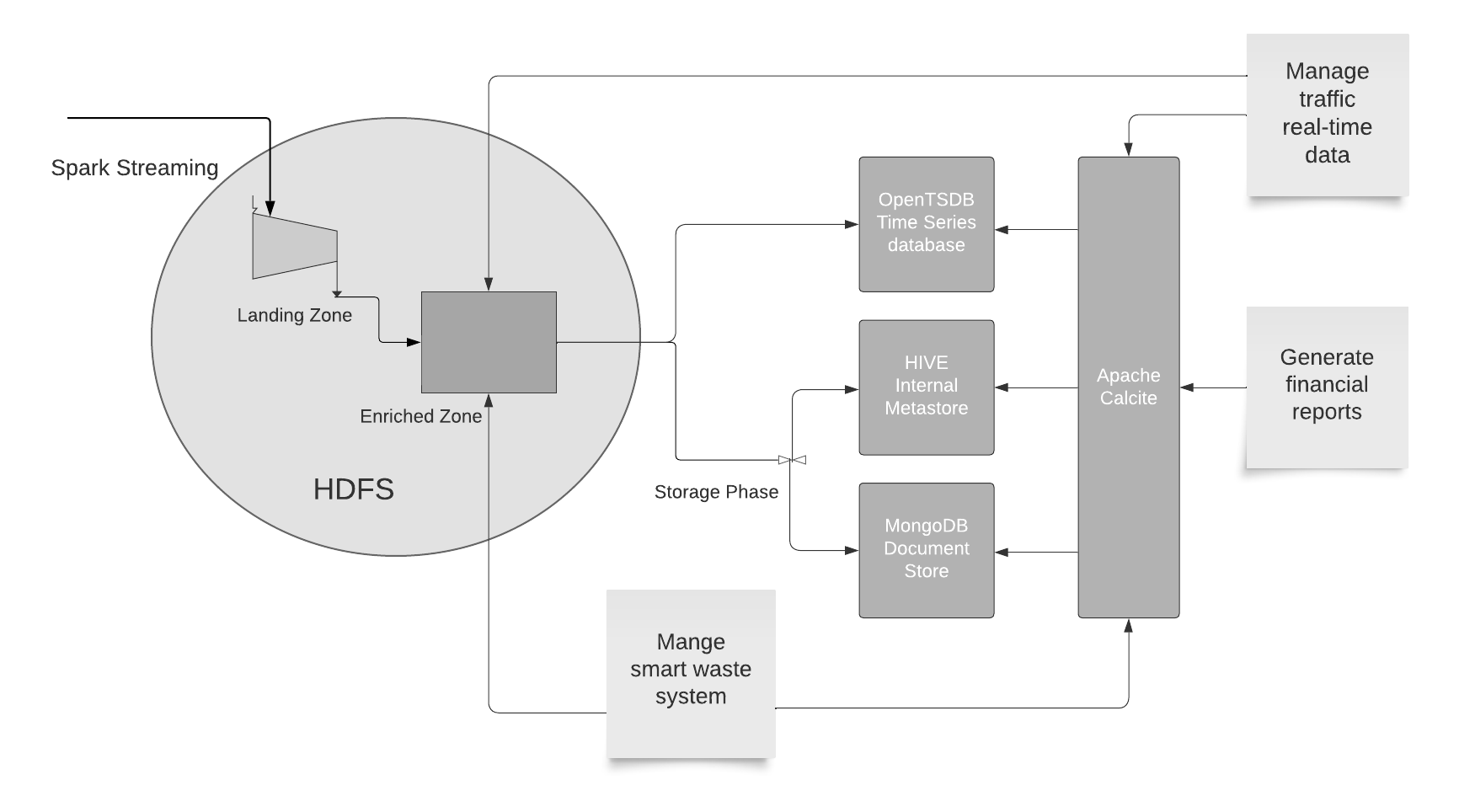}
\caption{Smart Cities Polyglot Data Processing}
\end{figure}

As shown in figure 11, the smart city initiative uses Apache Calcite to create a unified view of its heterogeneous data landscape, allowing city analysts to perform comprehensive analyses using familiar SQL queries. At the same time, Apache Spark processes real-time data streams to provide immediate insights and enable proactive management of city services. Together, these technologies empower the smart city to harness its diverse data sources effectively, turning raw data into actionable intelligence that enhances urban efficiency, sustainability and quality of life for its residents.

\section{Related Work}
Over recent years, numerous researches have been undertaken in the processing of social network data through the utilization of the Hadoop ecosystem, with the goal of gathering, storing and analyzing data to perform tasks such as sentiment analysis and misinformation investigation. By contrast, research in the field of polyglot data processing using the Hadoop ecosystem is an emerging area that addresses the complexities and challenges of managing diverse data types in large-scale environments and harnessing the power of multiple programming languages, computational paradigms, frameworks and tools within the Hadoop-Spark ecosystem to optimize data processing tasks. 

In [20], the authors explore the concept of polyglot persistence, an approach increasingly recognized for its effectiveness in data management. As detailed in their work, polyglot persistence is strategically designed to combine the advantages of various data stores while avoiding their respective limitations. The article offers a comprehensive overview of polyglot persistence tools, such as Polybase, and delves into related systems like Apache Calcite, providing a thorough summary of these advanced data management solutions.

Several works use the MapReduce programming model to implement social media data analysis. [54] proposes the use of Hadoop for processing Twitter data and MapReduce for sentence analysis, text mining, and multi-label classification. [51] uses the Twitter streaming API to collect data and MapReduce to perform sentiment analysis over the collected data. In [40], the authors posit that utilizing JSON files with Hadoop offers benefits in that information is stored in a key-value format, which in turn is used as input by MapReduce.

As [61] states, the design of data structure for social network analysis should be based on Hadoop massive datasets interface to meet the requirements of data processing under distributed development environment. The authors use MapReduce for raw data processing and iterative calculation of PageRank value. And in a comparison of data processing tools in Hadoop, such as MapReduce, Hive, and Pig, [45] conclude that, when it comes to unstructured data, MapReduce proves to be the most efficient tool.

The prevalent approach employed in the related studies to accomplish their objectives is the utilization of the MapReduce paradigm. However, there exists a noticeable gap within the existing body of literature concerning the utilization of Apache Spark, an alternative distributed computing framework, conceived as a response to the performance limitations inherent in the MapReduce paradigm and holding particular significance in the domain of big data analytics, particularly when dealing with large-scale datasets such as those found in social networks.

According to [5], the adoption of in-memory buffers as a replacement for intermediate disk files is what contributes to Spark's superior speed in comparison to Hadoop MapReduce. Indeed, Spark is particularly well-suited for processing large datasets. Moreover, [17] state that Spark mitigates the need for frequent read and write cycles, resulting in a tenfold improvement in performance compared to Hadoop when processing applications on disk. Additionally, the retention of intermediate data in memory renders Spark a hundredfold faster in memory-intensive scenarios.

Beyond the foundational works associated with the Hadoop ecosystem and the integration of Apache Spark, an emerging field of research is developing, centered on the concept of polyglot data processing. This area of study is dedicated to exploring the integration of multiple processing and storage components, aiming to provide efficient big data solutions, particularly for complex tasks like social network analysis. In parallel, the field is also investigating the strategic use of different storage systems, ranging from HDFS for handling massive datasets to NoSQL databases like HBase, which offer more agile management of unstructured and semi-structured data.
\section{Conclusions and Future Work}
In conclusion, the exploration of Polyglot Big Data Processing within the Hadoop ecosystem has underscored the ecosystem's versatility and robustness in addressing a broad spectrum of data processing requirements and scenarios. By leveraging a polyglot approach, organizations can tailor their data architecture to utilize the most appropriate data stores and computing frameworks, thereby optimizing performance, scalability and efficiency. The Hadoop ecosystem, enriched with a variety of data stores such as HBase, Hive and integrations with systems like Neo4J, offers a comprehensive platform that can accommodate the diverse nature of big data. The adaptability of Hadoop to function seamlessly with these varied data stores, coupled with its innate ability to process massive datasets, positions it as a cornerstone in the realm of big data processing.

Furthermore, the inclusion of computing frameworks like Apache Spark and Apache Storm within the Hadoop ecosystem enhances its capability to not only batch process vast amounts of data but also to perform real-time analytics and stream processing. This flexibility ensures that whether the task at hand involves analyzing historical data or processing real-time data streams, the Hadoop ecosystem is equipped to provide insightful, timely and actionable results. Through the use cases presented, ranging from healthcare analytics to social media analysis and financial market monitoring, it's evident that a polyglot big data processing approach, underpinned by the Hadoop ecosystem, offers a potent solution to the complex challenges presented by the multifaceted nature of big data. This approach not only facilitates the efficient and effective analysis of data but also enables organizations to harness the full potential of their data assets, driving innovation, enhancing decision-making and fostering a competitive edge in the data-driven landscape.

Future work should include executing comprehensive tests across the various domains cited in the article, such as healthcare, social media and financial markets. These tests will help in understanding the practical implications of implementing polyglot big data processing within these sectors, focusing on metrics like performance and scalability. Evaluating the Hadoop ecosystem's adaptability and efficiency in real-world scenarios will provide valuable insights into its application in diverse industry contexts. Another crucial area of future research is the comparative analysis of different polystores within the Hadoop ecosystem. By systematically comparing the performance, ease of use and integration capabilities of various polystores, researchers can identify optimal strategies for data management and processing in polyglot environments. This analysis will help organizations make informed decisions when architecting their big data solutions, ensuring they leverage the most suitable polystore options for their specific needs. Investigating the use of mediators to retrieve data from different data stores is essential for enhancing the efficiency of polyglot data architectures. Future work should explore the development and optimization of mediator systems that can seamlessly integrate with the Hadoop ecosystem, providing a unified query interface across diverse data models and stores. This research will contribute to simplifying data access and manipulation, enabling more flexible and powerful data analytics solutions. 

Innovation in the field of polyglot big data processing is pivotal for advancing the capabilities and applications of big data technologies. Future developments should focus on creating more cohesive and user-friendly approaches to managing and processing data across varied data models within the Hadoop ecosystem. Efforts could include enhancing interoperability between different data stores, improving real-time data processing capabilities and developing advanced analytics tools that provide deeper insights. Innovations that offer a unified view of different data models and facilitate more efficient data integration and analysis will drive the next wave of advancements in big data processing, empowering organizations to unlock new levels of value from their data assets. The journey toward more integrated, efficient and versatile polyglot big data processing systems is ongoing. By addressing these areas of future work, the field can continue to evolve, offering increasingly sophisticated tools and methodologies that cater to the complex and dynamic nature of modern data landscapes.

\end{document}